\documentclass[conference,compsoc]{IEEEtran}
\usepackage{graphicx}
\usepackage{subfigure}
\usepackage{float}
\usepackage{verbatim}
\usepackage{multirow}

\ifCLASSOPTIONcompsoc
  \usepackage[nocompress]{cite}
\else
  \usepackage{cite}
\fi
\ifCLASSINFOpdf
\else
\fi

\usepackage{xcolor}
\usepackage{xspace}

\newcommand{\ps}           {{PowerStack}\xspace}
\newcommand{\ct}           {{co-tuning}\xspace}
\newcommand{\at}           {{auto-tuning}\xspace}

\makeatletter
\newcommand{\linebreakand}{%
  \end{@IEEEauthorhalign}
  \hfill\mbox{}\par
  \mbox{}\hfill\begin{@IEEEauthorhalign}
}
\makeatother

\begin{document}
\title{Toward an End-to-End Auto-tuning Framework in HPC PowerStack}
%


\author{\IEEEauthorblockN{Xingfu Wu}
\IEEEauthorblockA{Argonne National Laboratory\\ The University of Chicago, USA \\
Email: xingfu.wu@anl.gov}
\and
\IEEEauthorblockN{Aniruddha Marathe}
\IEEEauthorblockA{Lawrence Livermore National Laboratory\\ Livermore, CA, USA \\
            Email: marathe1@llnl.gov}
\and
\IEEEauthorblockN{Siddhartha Jana}
\IEEEauthorblockA{Intel Corp., USA  \\
            Email: siddhartha.jana@intel.com} 
\and
\IEEEauthorblockN{Ondrej Vysocky} 
\IEEEauthorblockA{IT4Innovations National Supercomputing Center\\  Czech Republic\\
            Email: ondrej.vysocky@vsb.cz}
\and
\IEEEauthorblockN{Jophin John}
\IEEEauthorblockA{Technical University of Munich\\ Munich, Germany\\
            Email: john@in.tum.de}
\and
\IEEEauthorblockN{Andrea Bartolini} 
\IEEEauthorblockA{University of Bologna\\  Bologna, Italy \\
            Email: a.bartolini@unibo.it}
\and
\IEEEauthorblockN{Lubomir Riha} 
\IEEEauthorblockA{IT4Innovations National Supercomputing Center\\  Czech Republic \\
            Email: lubomir.riha@vsb.cz}
\and
\IEEEauthorblockN{Michael Gerndt}
\IEEEauthorblockA{Technical University of Munich\\ Munich, Germany\\
            Email: gerndt@in.tum.de}
\and
\IEEEauthorblockN{Valerie Taylor}
\IEEEauthorblockA{Argonne National Laboratory\\ The University of Chicago, USA \\
Email: vtaylor@anl.gov}
\linebreakand 
\IEEEauthorblockN{Sridutt Bhalachandra}
\IEEEauthorblockA{Lawrence Berkeley National Laboratory, USA \\ 
Email: sriduttb@lbl.gov}
}


\maketitle

\thispagestyle{plain}
\pagestyle{plain}

\begin{abstract}

Efficiently utilizing procured power and optimizing performance of scientific applications under power and energy constraints are challenging. The HPC PowerStack defines a software stack to manage power and energy of high-performance computing systems and standardizes the interfaces between different components of the stack.
This survey paper presents the findings of a working group focused on the end-to-end tuning of the PowerStack. First, we provide a background on the PowerStack layer-specific tuning efforts in terms of their high-level objectives, the constraints and optimization goals, layer-specific telemetry, and control parameters, and we list the existing software solutions that address those challenges. Second, we propose the PowerStack end-to-end auto-tuning framework, identify the opportunities in co-tuning different layers in the PowerStack, and present specific use cases and solutions. Third, we discuss the research opportunities and challenges for collective auto-tuning of two or more management layers (or domains) in the PowerStack. This paper takes the first steps in identifying and aggregating the important R\&D challenges in streamlining the optimization efforts across the layers of the PowerStack.

\end{abstract}


\section{Introduction}

As we enter the exascale computing era, power and energy management are key design points and constraints for any next generation of supercomputers \cite{BB20}. Efficiently utilizing procured power and optimizing the performance of scientific applications under power and energy constraints are challenging for several reasons including dynamic phase behavior, manufacturing variation, and increasing system-level heterogeneity. While  several individual techniques have been proposed for the automatic and efficient management of power and energy, the majority of these techniques have been devised to meet the needs of a specific high-performance computing (HPC) center or specific optimization goals. Specifications such as PowerAPI \cite{GL16, GR19}, IPMI \cite{3}, and Redfish \cite{redfish} provide high-level power management interfaces for accessing power knobs. A recent survey \cite{MA18} conducted by the EEHPC WG \cite{5} concluded that the majority of such techniques lack the application-awareness required to achieve the best system performance and throughput. Furthermore, each technique tends to improve the management of power and energy for a different subset of the site or system hardware and at different (and often conflicting) granularities. Unfortunately, the existing techniques have not been designed to coexist simultaneously on one site and cooperate on management in a streamlined fashion. 

To address these gaps, the HPC community needs a holistic stack for power and energy management. The HPC \ps \cite{4, BB20} started in May 2018 as a working group to gather the experience of active developers in industry, computing centers, and academia in building software interfaces and solutions for handling and optimizing the power and energy consumption in HPC systems in production. \ps defines a software stack that manages the power and energy of HPC systems and standardizes the interfaces between different levels of software components in the stack. One of the key aspects of \ps is to define a vision of a holistic power and energy management stack extensible by design and capable of optimizing the target power- or energy-efficiency application-aware metric so that it can trade off power, energy, and time to solution in order to optimize the efficiency of an HPC application. Its second aspect is to define a standard interface to interact with optimization software and hardware knobs across different vendor HPC systems. 
Based on the state of the art of the components available in the community for power and energy management \cite{GL16, GR19, 3, redfish, MA18, 5}, a hierarchical straw man \ps design \cite{1, BB20} was proposed to manage power and energy at three levels of granularity: the system level, the job level, and the node level. This implies the need to put in place the following incrementally: 

\begin{itemize}
    \item Define a site-level requirement, a power-aware system Resource Manager (RM) / job scheduler, a power-aware job-level manager, and a power-aware node manager.
    \item Define the interfaces between these layers to translate objectives at each layer into actionable items at the adjacent lower layer.
    \item Drive end-to-end optimization across different layers of the \ps.
\end{itemize}

To address this need, we formed an \ps End-to-End Auto-tuning Working Group in 2019. 
A plethora of literature on power-aware tuning exists, including notable works by the members of this working group. These efforts include system and hardware tuning, tuning with runtime system, application-level tuning, compiler-level parameter tuning, loop-level parameter tuning, and deep learning-based hyperparameter tuning.

A primary limitation of most---if not all---of these efforts is that the tuning research has been solely limited to the individual layers of the \ps. The opportunities for further gains in power efficiency from collectively tuning two or more layers of the \ps have largely remained untapped. The goal of this paper is to explore those untapped opportunities by addressing the following specific questions. 
 
 \begin{itemize}
\item	\textbf{Opportunity analysis}: How do we quantify the potential benefits of end-to-end \at across the different layers of the \ps? What experimentation is required to achieve baseline quantification of the benefits of end-to-end \at?
\item	\textbf{Identification of high-level challenges}: What are the high-level research questions to be explored in the end-to-end \at of the \ps? What engineering solutions and research approaches are needed to these questions?
\item	\textbf{Interaction of existing layer-specific tuning}: Based on the conceptual diagram of the \ps, what interactions are required across the layers of the \ps with existing layer-specific tuning approaches as a precursor to end-to-end \at? 
\item	\textbf{Extension toward end-to-end \at}: How do we combine the existing \at approaches to develop comprehensive end-to-end \at solutions for the high-level power and energy goals? 
\end{itemize}

To our knowledge, this paper is the first attempt at identifying and aggregating the important R\&D challenges in streamlining the interoperation across the layers of the \ps. We present the important high-level questions, concrete ideas, and ongoing efforts discussed by the members of the \ps working group. The remainder of this paper is organized as follows. Section 2 surveys the  layer-specific tuning efforts in terms of their high-level objectives; describes the high-level objectives; discusses the target metrics, layer-specific control parameters, and specific methods; and lists the existing software components. Section 3 proposes the \ps end-to-end \at framework, identifies the opportunities in collective tuning (henceforth \textit{co-tuning}) different layers in the \ps, and presents the specific use cases. Section 4 identifies the further opportunities and open challenges for \ct of two or more management layers (or domains) in the \ps. Section 5 summarizes our conclusions.

\section{Survey of PowerStack Layer-Specific Tuning}


In this section, we first describe the high-level objectives of the existing layer-specific tuning approaches at the different layers of the \ps: system (i.e., cluster), job / application, and node. Next, we outline the target metrics for the existing tuning approaches. We then present the layer-specific control parameters, telemetry and specific methods used to accomplish the objectives by the individual layers.

\subsection{Objectives}

The common objective of each layer of the \ps is to operate within the power constraints or energy goals assigned by the upper layer. A power constraint is applied and measured over a time window. An energy goal is assigned and measured over the job execution or system uptime. The smallest supported time window is defined by what can be supported at each layer and what is acceptable by the upper layer. Along with the primary objective of adhering to a power constraint, the following secondary metrics are targeted:
\begin{itemize}
\item	Power-constrained performance optimization
\item	Performance constrained energy optimization, RM-brokered SLA-compliant performance
\item	Guaranteed rate of change, or lower and upper bounds on power (power usage) in system in a specified time window
\item	Thermal-constrained performance optimization
\end{itemize}

Some of the secondary metrics monitored and affected in this process are as follows:
\begin{itemize}
\item	System utilization, resource utilization
\item	Thermal metrics: ambient temperature, water temperature 
\item	Job turnaround time,  queuing delay, throughput
\item	Reduced memory footprint, reduced data movement and I/O footprint
\end{itemize}

\subsection{Metrics}

The objectives at different layers of the \ps can be realized by using measured or derived metrics at those layers as follows: 

\begin{itemize}
\item	Job-level power (watts) or energy (watt hour or joules) usage, 
\item	Execution time (seconds/minutes/hours)
\item	Operating frequency (Hz)
\item	Performance (FLOPS, IPC, IPS)
\item	Power efficiency (FLOPS/watt, IPC/watt)
\item	Energy efficiency ($ED, ED^2, FLOPS/Joule, \\IPC/Joule$)
\item	Node utilization (\% of time in use, \% of resource in use)
\end{itemize}

\subsection{Methods and Parameters}

The objectives described above are realized by each layer by managing a set of available controls provided by the adjacent lower layer. The parameters are tuned through the available methods provided by the hardware or indirectly managed by the lower layers (runtime, node-level manager, system software, etc.). We describe the parameters and the methods used by the individual layers of the \ps at system, job/runtime, application, and node levels shown in Table \ref{tab:data}.

\begin{table}
\center
\caption{Survey of parameters and methods used by the layers of the \ps}
\begin{tabular}{c}
  \includegraphics[width=.45\textwidth]{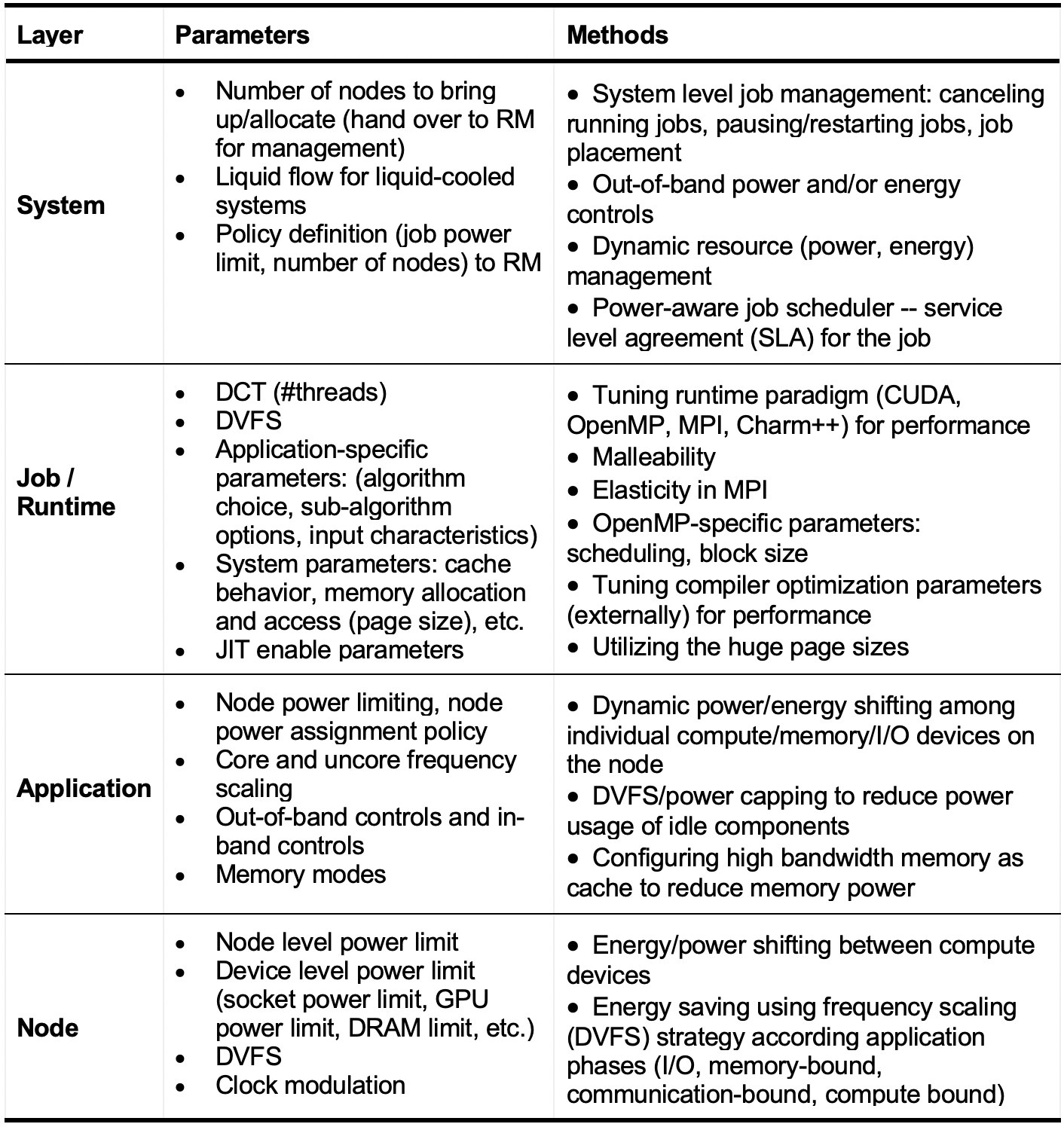}
  \end{tabular}
\label{tab:data}       
\end{table}  

\subsection{List of Existing \ps Components}

In Table \ref{tab:tl} we list several existing software components at the four layers: resource manager/job scheduler, job-level runtime system, node-level management, and application-level tuning. By integrating proper software components from each layer, we are able to do the proposed end-to-end \at in \ps.

\begin{table}
\center
\caption{Existing tools/solutions at each layer of the \ps}
\begin{tabular}{c}
  \includegraphics[width=.45\textwidth]{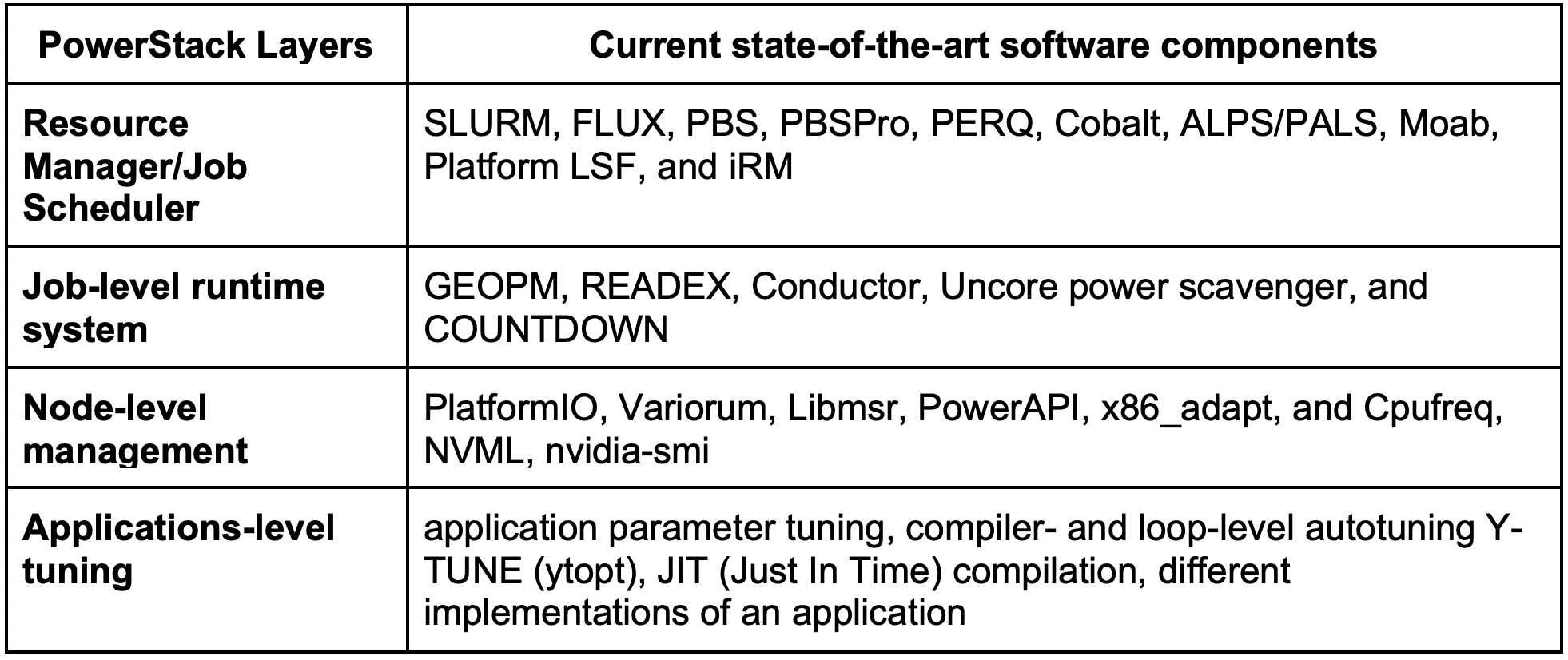}
  \end{tabular}
\label{tab:tl}       
\end{table}

\section{End-to-End Auto-Tuning Framework}

Before we delve into \textit{end-to-end tuning} of the \ps, we define the term \textit{tuning}. At a high level, \textit{tuning} is the process of improving the target metric through better  handling of available control parameters and configuration options without violating operating constraints (if any). The process of tuning in the layers of the \ps (a) typically targets performance or power efficiency as the primary metric, (b) complies with the operating power constraint imposed on the layer, and (c) attempts to improve the management and orchestration of the available control parameters that affect the application and/or hardware performance. In this process, the other layers are either treated as black boxes or are ignored altogether in order to keep the research problem tractable. Extending this definition, we define \textit{co-tuning} as the process of improving the target metrics of two or more layers of the \ps by incorporating cross-layer characteristics in the orchestration process. End-to-end \at aims to perform holistic \ct of all layers of the \ps. 

\begin{figure}
\center
 \includegraphics[width=.45\textwidth]{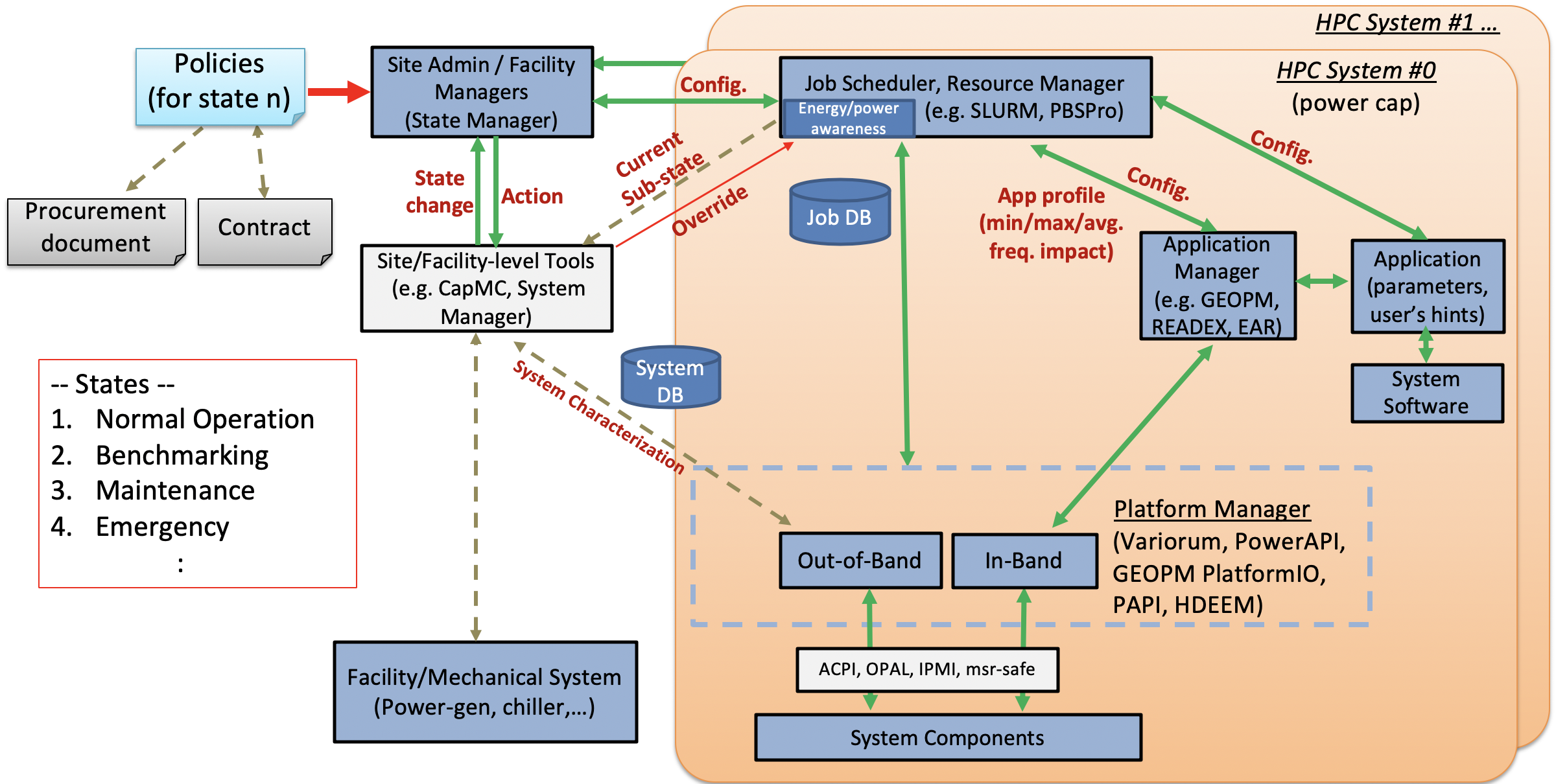}
 \caption{High-level overview of end-to-end \at framework (orange boxed portion)}
\label{fig:1}
\end{figure}

In 2019, the \ps community sketched a schematic diagram outlining the different components of a power management stack, shown in Figure \ref{fig:1}.  A site has one or more HPC systems, site policies, and a power budget. Each system is constrained under a derived system-level power budget. For end-to-end \at in \ps, we will focus on tuning at the system, job/application, and node. We propose a high-level overview of the end-to-end \at framework (orange boxed portion) in Figure \ref{fig:1}. We describe the knobs at each layer, what control knobs can be modified on temporal and spatial dimensions, who controls the knobs (actors), and what metrics can be measured. We define tunable parameters at each layer, then discuss how to \at the combination of different parameters at the distinct layers (parameter space) for an optimal solution (the smallest runtime, the lowest power, or the lowest energy) under a system power cap. 

The traditional \ps design has focused  largely on the engineering challenges in standardization and deployment. In contrast, this paper focuses on extending the current design of the \ps to address novel research challenges in end-to-end \at of the \ps. Specifically, we extend the traditional \ps model by considering two additional, largely static layers: 1) Application: We consider application as its own \at layer; and 2) System software: We consider the system software such as the compiler toolchain, system-level dependencies such as MPI, OpenMP, and thread-management libraries, and other external entities that play an important role in realizing the \ps but have no direct interfaces in the traditional design.

\subsection{Opportunity Analysis}

In this section, we survey the opportunities in collective \at of two or more management layers in the \ps. The goal of this survey is to find potential areas for research and prepare a list of broad research questions that the \ps is collectively interested in tackling. An outcome of this survey would be to come up with research activities that the \ps community can collaboratively participate in depending on the area of expertise.

\begin{figure}
\center
 \includegraphics[width=.35\textwidth]{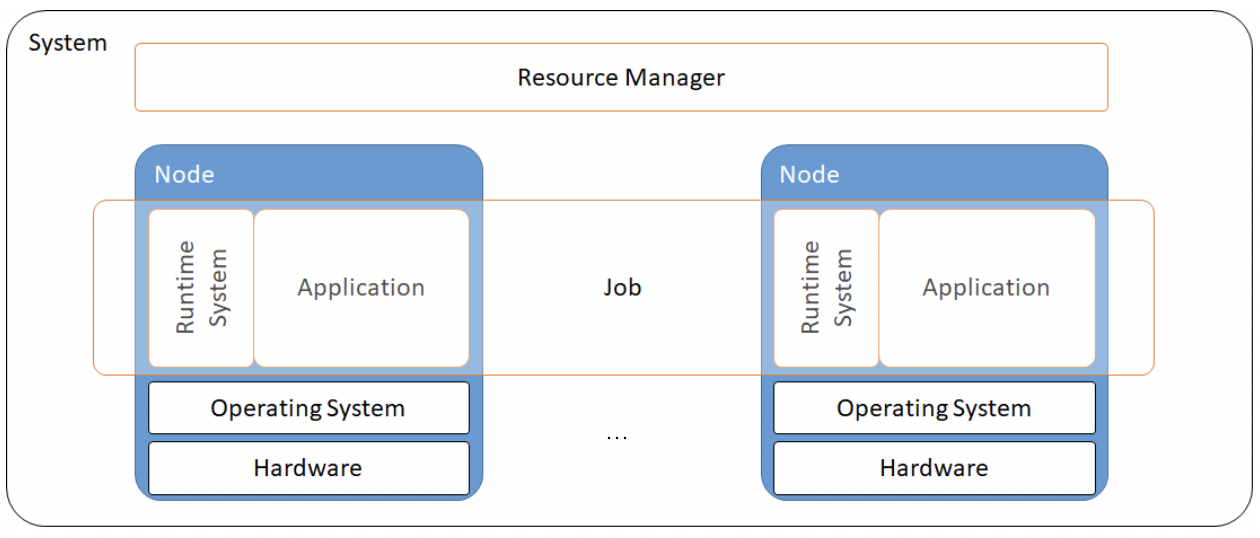}
 \caption{Placement of Resource Manager, Job, Runtime System, and Application (orange lines)}
\label{fig:2}
\end{figure}

Before we discuss the \ct opportunities for individual layers, we define the important terms used in the rest of this paper. These terms are listed in Table \ref{tab:df}.  Figure \ref{fig:2} shows a high-level overview of the placement of Resource Manager, Job, Runtime System, and Application in the \ps, and the interaction between the layers (orange and green lines). 

\begin{table}
\center
\caption{Definitions of terms}
\begin{tabular}{c}
  \includegraphics[width=.45\textwidth]{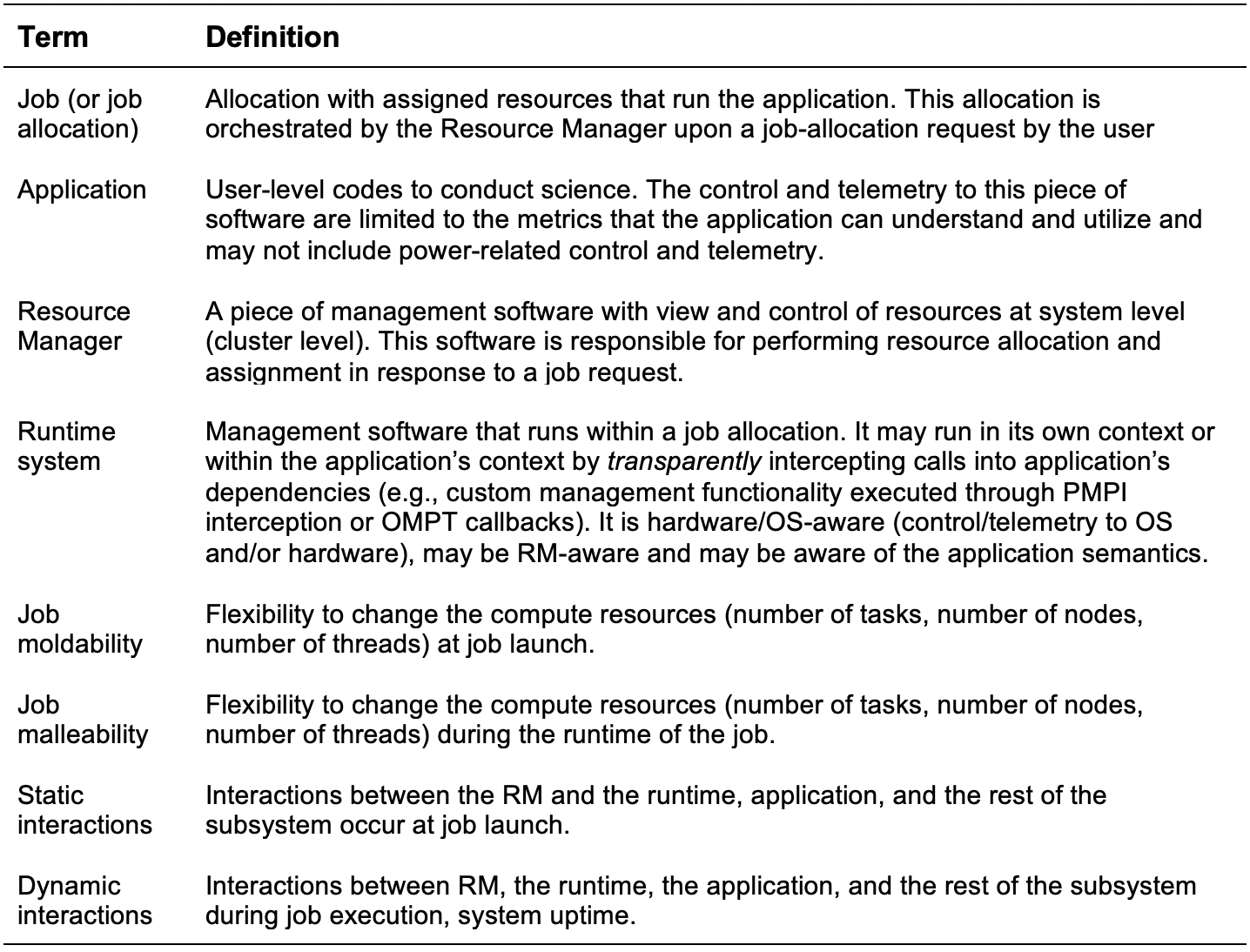}
  \end{tabular}
\label{tab:df}       
\end{table}  

\subsubsection{Co-Tuning Resource Manager and Runtime System}
\par \noindent The objective of this \ct is to explore and leverage opportunities in the simultaneous tuning of the resource manager with empirical or online knowledge of the dynamic behavior of a power-aware runtime system subject to a target power constraint or energy efficiency metric. We consider two directions of interaction in this section---from resource manager (RM) to runtime system and vice versa. Several types of interactions between the RM and the runtime system can occur, as outlined below.

\hspace{0pt} \par \noindent \textbf{Static interaction}: These interactions define the management decisions by the resource manager at job launch.
\begin{itemize}
\item How many nodes (for moldable jobs). The user provides a minimum and a maximum number of nodes the job can use.
\item	Which nodes (or compute resources) to select for job launch for managing inefficiencies in the system such as thermal hot spots, and processor manufacturing variation.
\item	Which job to run (or backfill) from the job queue.
\item	Which binary dependencies to pick given the situation on the cluster.
\end{itemize}

\par \noindent \textbf{Dynamic interaction}: These interactions define the management decisions by the resource manager during job runtime.
\begin{itemize}
\item	How much power to reassign to a running job (reduce or increase).
\item	Which job to pause (run queue) or restart (wait queue) if supported by the job.
\item	Whether the resource manager or the runtime system can leverage job malleability. Some resource managers and runtime systems already leverage malleability at the thread level through concurrency throttling. 
\item	Job relaunch. Some resource managers may explore just-in-time (JIT) compilation of the application to relaunch the job either through checkpoint-restart in the same job or pause/resume over different allocations.
\end{itemize}

Another type of categorization of the interactions between the resource manager and the runtime system is based on the job awareness.
\begin{itemize}
\item \textbf{Job-aware interactions}: These are the interactions between the resource manager and runtime system that take job behavior into account when applying power management decisions. The job awareness is based on either the empirical profile of the application or runtime telemetry collected from the application.
\item \textbf{Job-agnostic interactions}: These interactions between the resource manager and runtime system do not take job behavior into account. These are primarily the interactions  agnostic (or transparent) to the application itself.
\end{itemize}

The interaction from the resource manager to the runtime system may occur through reassignment of power controls and reporting of degradation in performance or efficiency observed at the system level with a heartbeat signal.
The interaction from the runtime system to the resource manager may occur through reporting of job-level power usage, request for additional power usage or returning unused power, and non-power-related controls that indirectly affect job and system power efficiency.
The	potential metrics by which the opportunity can be measured are system-level energy efficiency and system-level job throughput.

\subsubsection{Co-Tuning Resource Manager, Application, and System Software} 
\hspace{0 pt}\\ The objectives for the co-tuning are to assign resources to the job, control their consumption,  select heuristics to maximize power efficiency at the system level, and select optimal application control parameters at application launch. Note that the job-level runtime is either absent or agnostic to the \ct of the resource manager, the application, and the system software.

The goal is to understand what opportunities exist for tuning two interactions in this \ct space: (a) tuning the control loop in RM by making it application-characteristics aware through empirical data or on-line monitoring and (b) requesting changes to resource assignments by the application for further tuning control parameters within the application. The objective is to maximize the performance of all applications on the system under a system-wide power limit by maximizing per-job power efficiency (minimum performance impact).

We assume that in this interaction, the control and telemetry information to and from the application is limited to application-centric data. This does not include power-related controls and telemetry as noted in the definitions, since the application itself is not involved in power management decisions. For the interaction from the resource manager to application, the following are involved: 1) power and/or energy budget if the application understands the metric;  2)  power efficiency translated into an application-level metric such as watts per timestep (But this will be application-specific and will not scale for all applications); 3) number of resources; 4) other operating constraints: from job to managed resource;  5) power/frequency/concurrency control; 6) application control parameters; and 7) application launch parameters. Interaction from the application to the resource manager may occur in terms of reporting metrics describing application progress. The required interfaces between the resource manager and the application include power consumption monitoring, power limit specification, the expected efficiency metric to monitor from job-level runtime to RM, and the expected efficiency metric to monitor from application to job-level runtime.

\subsubsection{Co-Tuning Runtime System and Application, System Software}
\hspace{0 pt}\\ The objective is to fine-tune system parameters, software, and application parameters at job launch as well as runtime in order to maximize job power efficiency under the power budget. The runtime system may leverage JIT compilation of the application with static actions such as the amount of required computational resources (\#threads, \#processes), affinity and binding, and JIT enable parameters  and with  dynamic actions such as dynamically adjusting the amount of computational resources, affinity, and binding, turning on/off JIT enable parameters for application components, power limit (+ window size) specification, and dynamic power management.

The required interfaces are power consumption monitoring, CPU power management interfaces (power, frequency tuning),  access to specific registers (not only) for uncore frequency tuning, and hardware performance counter monitoring. The input and output variables/dependencies for doing power management at these layers are 
the input variables are CPU power limit, dynamic power management, number of resources, thermal design power (TDP), and the required configuration that should be applied during  runtime.
The output variables are to confirm/deny the configuration set assigned by the runtime system, power and energy consumption, and performance in terms of application-centric metrics. The potential metrics to define the opportunity are high performance under power cap, low energy under power cap, and full system utilization.

\subsubsection{Co-Tuning Resource Manager, Runtime System,  Application, and System Software}

\hspace{0 pt}\\ The objective of this \ct space is to explore opportunities for how the resource manager, runtime system, and  application (along with system software) can be co-tuned to maximize application performance under a power constraint. 
The interactions are the required interfaces for interaction across three layers. The interfaces must be defined to answer the following questions:
1)	What interfaces are needed to translate high-level targets at the resource manager into targets for the job-level runtime system and the application? and
2) What telemetry interface needs to be provided from application to the job runtime and from the runtime system to the resource manager? The discussions of the pairwise \ct process described previously cover the interfaces required in this \ct process.

For example, a target metric of throughput under a system-level power constraint at the resource manager level needs to be translated into power efficiency targets or total runtimes of individual jobs managed by the job-level runtime system subject to a job-level power constraint. This must be translated into improvements in the calculations per simulation step per watt at the application level.


\subsection{Specific Use Cases and Solutions}

In this section, we discuss seven specific use cases and solutions for collective tuning of the resource manager, runtime, application, and node layers. 

\subsubsection{Co-tuning of Resource Manager (SLURM), Runtime System (Conductor), and Application (Hypre Library)}

 This use case describes the \ct of the resource manager, runtime system, and application. The target application is a 27-point Laplacian problem implemented as part of the test program shipped with the Hypre linear solver library \cite{HYPRE02}. The control parameters exposed by Hypre are primarily at the algorithm and subalgorithmic level. Specifically, Hypre enables the user (or the runtime system) to select input preconditioner, linear solver, subsolver options, and postconditioner. In our experience,  several thousand combinations of these options  can be selected from at job launch. While tuning Hypre parameters has been an extensively researched topic, our empirical analysis showed that subjecting system-level power constraints severely affects the applicability of previous tuning efforts \cite{AR17}.  We observed that the best-case combination of the tuning knobs for Hypre is often inefficient when subject to a hardware power constraint. Consequently, tuning the resource manager and runtime system without taking into account how the power-aware runtime system and Hypre collectively respond to the decision layers leaves performance on the table. We use the Conductor runtime system \cite{AP15} to transparently optimize the job-level power budget on the allocated nodes. Conductor exposes control parameters that impact the granularity and efficiency of its power-balancing algorithm under the assigned job-level power limit.


Our target metrics are twofold:  1) improve power efficiency in terms of instructions per cycle (IPC) per watt at the runtime system level. and 2) improve job throughput at the resource manager level (\#jobs/hour). This presents a multidimensional optimization space at the resource manager level that leads to the following research challenge: How does the resource manager select efficient control parameters at job launch while incorporating the impact on overall job throughput? We plan to explore interfaces to specify runtime system and Hypre parameters that can  be selected only at job launch. During the application execution, we plan to explore runtime and subalgorithm options that can be varied.

At the application level, we will explore several parameters including the choice of solver algorithm, subsolver options, and data preconditioner (smoother and coarsening). At the runtime system level, we explore traditional platform knobs such as power limiting and frequency scaling. At the resource manager level, we explore the number of nodes and MPI tasks as the control parameters. The approach relies on the telemetry at different layers of the \ps. The primary metrics for telemetry include power usage (watts), operating frequency (GHz), and IPC.
The primary challenges include managing static interactions between the layers at application launch and estimating the impact of several control parameters on the application behavior.

\subsubsection{Co-tuning of Resource Manager (SLURM) and Runtime System (GEOPM)}

\begin{figure}
\center
 \includegraphics[width=.4\textwidth]{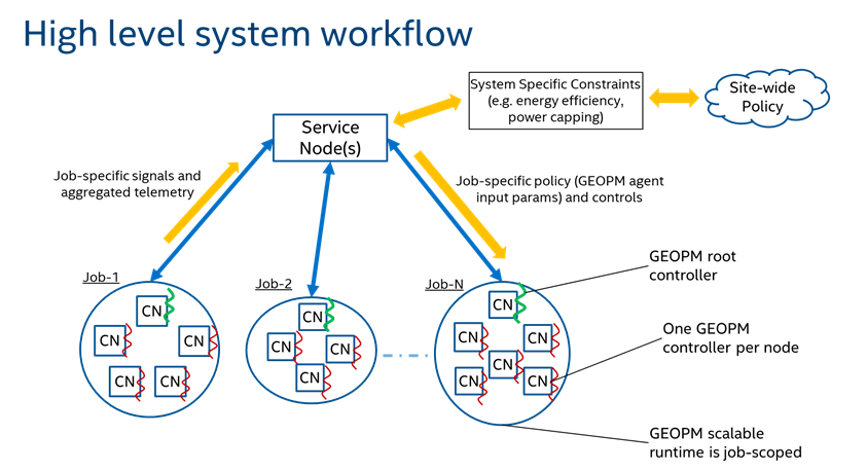}
 \caption{High-level overview of multijob GEOPM policy assignment}
\label{fig:4}
\end{figure}

One of the critical challenges at HPC facilities is to limit energy/power consumption of their in-house systems based on their contractual obligations with utility service providers \cite{5}. In order to enhance efficiency,  a system-level power-aware software agent (like the resource manager) is needed to frequently monitor and control its consumption in tandem with an application-level power management software agent (like GEOPM \cite{GEOPM, ES17}). Figure \ref{fig:4} illustrates how facility-level power policies filter down into job-level granularity.  

The community is working toward a prototype that facilitates a bidirectional communication channel for interoperation between the resource manager and GEOPM. The objectives are  systemwide characterization of frequency, power, and thermal variation across the system plus node outlier detection, node energy savings or FLOPS improvements via adapting CPU/GPU PM controls according to application phases or characteristics, overall application energy savings or runtime reductions via steering power between nodes according to load imbalance patterns, and enabling of other tools or resource managers to perform system-wide power/performance management by leveraging GEOPM.
	
GEOPM has three main modes of community site-level policies . 

1) Enforcing static preconfigured sitewide policy: This method relies on preconfigured policy configuration files passed  to GEOPM during job launch. These files can either be directly baked into the node boot image on a pseudo-file system or passed along to the job launcher. Adding extra awareness of these files within the RM enables it to enforce power management control with/without GEOPM.

2) Enforcing job-specific policies: This method is suitable for HPC sites that frequently run a finite set of applications with historic profile information about its power/energy/thermal characteristics. Such sites typically maintain a database that maps applications to specific policy parameters (e.g., fixed frequency or power-cap). 

3) Fully dynamic policy: This ongoing R\&D work relies on the entire power management stack participating in dynamic cooperation between the electric grid supply, the resource manager, and the job-level runtime (GEOPM). This mode enables job-specific hardware resources to react in accordance with the instantaneous system-level power/energy requirements realized at the site level.

Interfaces to system-level agents: The communication channel/interface can take the form of environment variables, preconfiguration files, job-launcher command line options, or shared memory. Having a shared memory facilitates efficient interaction and is referred to as an endpoint, which acts as a gateway between a persistent compute node daemon (like SLURM) and an application power-management daemon (like GEOPM root controller). 
GEOPM provides a plugin-based interface that enables users to plug-and-play their own algorithms of choice. By default, a typical GEOPM installation compares prepacked with five different algorithms that correspond to the most common policies among HPC sites: energy efficiency under a performance degradation threshold, power load balancing based on the average node power cap, static frequency assignment for the entire lifetime of the application, static power cap assignment for the entire lifetime of the application, and monitoring application energy/power metrics. The open design challenges are fault tolerance, security, multitenancy, interaction with other OS/user-level/RAS daemons, and handling of conflicting power management mechanisms on the same system.

\subsubsection{Co-Tuning Compiler (Clang), Application, and Runtime System: ytopt Auto-Tuning Framework}

\begin{figure}
\center
 \includegraphics[width=.4\textwidth]{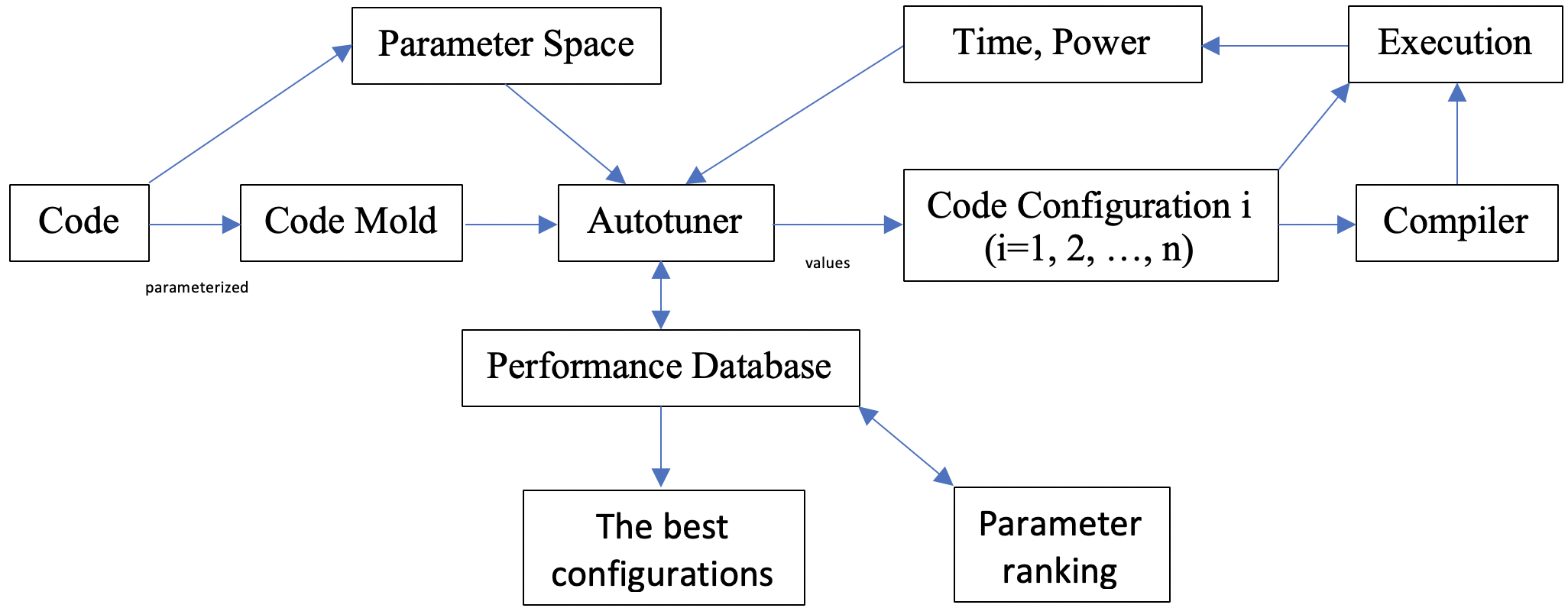}
 \caption{Autotuning application and runtime parameter selection}
\label{fig:7}
\end{figure}

In the \at \textit{ytopt} project\cite{WK20}, we use the new Clang loop tiling, interchange,  packing, and/or jam pragmas as examples to illustrate the integration process about \at the pragma parameters to achieve the optimal performance. Figure \ref{fig:7} shows our \at framework.

We analyze an application code to identify the important parameters that we  focus on, then replace these parameters (application parameters, loop transformation parameters) with  symbols such as \#P1, \#P2, ..., \#Pm in  top-down order to generate another code with these symbols as a mold code.  
We define the value ranges of these symbols for parameter space used by an autotuner. Three main steps are involved: 1) autotuner assigns the values in the allowed ranges (using random forests as default) and replaces these symbols in the mold code with them to generate a new code using the function plopper, 2) plopper compiles the code and executes it to get the execution time, and  3) autotuner outputs the time and the elapsed time with the parameters' values to a performance database. Steps 1, 2, and 3 are repeated until the maximum number of evaluation times set (using the option --max-evals=maximum number; default: 100) is reached. 
We process the performance database to find the smallest execution time and output the optimal configurations for the time. We also used counters-based performance, and power modeling to identify the most important counters to improve energy efficiency of HPC applications \cite{WU16}.

To apply this framework to the end-to-end \at involves parameter tuning for a combination of different parameters at the distinct layers (parameter space) under a system power cap. These parameters include  application parameters, system environment parameters such as setting number of threads, affinity, JIT-enabled parameters, power-capping size, and loop transformation parameters. We consider the metrics: runtime, power consumption, and energy consumption.
Under a system power cap, the framework can be used to find the best combination of different parameters for the optimal solution (the smallest runtime, the lowest power, or the lowest energy). The challenges and open issues are large parameter space for search and how to define tunable parameters at each layer for use.


\subsubsection{Co-Tuning of Runtime (READEX) and Application}

The Horizon 2020 project Runtime Exploitation of Application Dynamism for Energy efficient exascale computing (READEX)~\cite{READEX} came with an idea and tool suite providing a parallel application splitting into regions of different resource requirements and dynamic tuning of hardware parameters that suit the needs of each of the regions.

Besides the hardware and system parameters tuning, the READEX tool suite  supports static and dynamic tuning of the application parameters. For this purpose a static application configuration tuning (ACP) plugin and a dynamic application tuning parameters (ATP) plugin~\cite{PACOreadex} for the Periscope Tuning Framework~\cite{PTF} were developed. The static tuning requires the identification of parameters in the application's launch configuration file, which changes at the beginning of each run. The dynamic tuning requires the application instrumentation with the ATP's API to identify the tuned parameters. Each parameter's value is then changed at every $x$th iteration of a loop in which the instrumentation is, which requires support at the application side to handle such modification during runtime. The key input information for both static and dynamic application parameters tuning is not only a list of parameter values to set but also dependency conditions that express which combinations of parameters are not allowed.

\begin{figure}
\center
 \includegraphics[width=.4\textwidth]{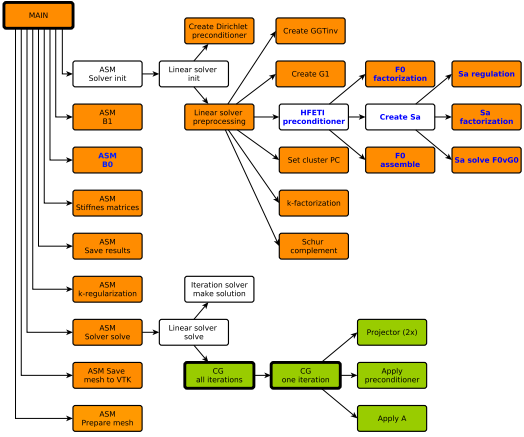}
 \caption{Graph of the ESPRESO FETI solver regions instrumented in the source code}
\label{fig:espreso}
\end{figure}

ESPRESO FETI solver~\cite{IJHPCAespreso} was tuned by using the READEX tools for optimal hardware configuration as well as using the  ATP to find an optimal solver, preconditioner, and domain size,  all of which have a major impact on the time to solution of the issued problem. The application has been instrumented with a set of regions as presented in Figure~\ref{fig:espreso}. Such optimization led to major runtime and energy savings as well as to improved scalability~\cite{IJHPCAespreso,PACOreadex}.


The complete READEX tool suite does the search of the optimal configuration automatically using one of many supported algorithms for the space state search, which could be potentially very large. However, the challenge of this approach is in the instrumentation of the application and its configuration file with valid and complete  dependency conditions for the application parameters.

\subsubsection{Co-Tuning Resource Manager (IRM) and Application Programming Model (EPOP)}
\begin{figure}
\center
 \includegraphics[width=.35\textwidth]{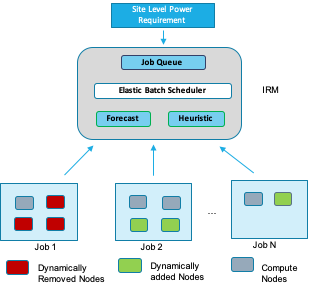}
 \caption{Dynamic resource redistribution to enforce power corridor}
\label{fig:irm}
\end{figure}

The site-level power budget (power corridor) is usually enforced by using job cancellation, idle node shutdown, power capping, and  dynamic voltage and frequency scaling. A new proactive power corridor management strategy is developed with our dynamic resource management infrastructure~\cite{tum1,tum2} comprising an Invasive Resource Manager (IRM) and Invasive Message Passing Interface (IMPI). The core research challenge is power management using dynamic resource redistribution among applications.

The power usages of running applications are predicted and analyzed for power budget violation. If a violation is predicted, steps are taken to formulate a new resource redistribution heuristic to satisfy the site-level power constraints. As shown in Figure~\ref{fig:irm}, the node distribution was dynamically changed by IRM to maintain the power budget. This dynamic strategy can also cater to dynamic power budgets arising because of renewable energy sources. We also constructed a programming paradigm called Elastic Phase Oriented Programming  (EPOP)~\cite{tum3} on top of IMPI for developing dynamic applications and for co-tuning. EPOP measures the power as well as performance characteristics of the application and communicates with IRM upon request. Using EPOP, the programmer can explicitly inform IRM u about the application phases where resource redistribution is needed or not. A dynamic resource manager also requires knowledge of application constraints (for example, the requirement of a cubic number of processes in LULESH) for efficient resource redistribution. Hence, a co-tuning among resource managers, runtime system, and applications is necessary in order to exploit the dynamism. Lack of running dynamic applications and application-specific constraints can hinder the calculation of successful resource redistribution. Power-aware scheduling with other strategies such as power capping and DVFS can address this shortcoming.

\subsubsection{Co-tuning Resource Manager (SLURM) and Runtime (COUNTDOWN)} 

COUNTDOWN is a runtime library for performance-neutral energy saving in MPI applications developed by the University of Bologna and CINECA \cite{CNTD,CNTDSLACK}. The energy saving is obtained transparently to the user, without requiring application code modifications or  recompilation of the application. 
The target metric is energy efficiency with no performance degradation seamlessly by leveraging MPI communication phases. We consider the metrics: runtime, power consumption, and energy consumption. The control parameters/units are the types and sections of the MPI calls in which to reduce the power consumption and CPU performance. The COUNTDOWN runtime will intercept those and separate waiting time from copy time and computing time. The COUNTDOWN configuration can be set at the beginning of a job run to (i)  profile only the MPI communication regions or (ii) reduce the power consumption during MPI wait and copy time or (iii) reduce the power consumption during MPI wait time only. The power reduction is achieved by interacting with the node level. While at the system level, the resource manager interacts with the COUNTDOWN configuration to select the level of aggressiveness (i.e., energy-saving). This enables co-tuning of application and communication phases power management.

The challenges and open issues are the dynamic adaptation of the COUNTDOWN configuration during job execution, its integration with the job scheduler, and the estimation during job execution of the accumulated time-to-solution overhead. 

\subsubsection{Co-tuning of Two Runtime Systems: COUNTDOWN and MERIC}

This use case for using two runtime systems at the same time optimizes the hardware parameters for the same metric. 
The COUNTDOWN is a powerful tool for optimizing the communication phases of the MPI applications. However, it does not take into account  other application characteristics such as I/O and memory-bound as well as compute-bound regions of the code that may provide a major space for savings exploitation~\cite{READEXbook}.
The MERIC library~\cite{MERIC} implements the READEX approach, so it tunes the application based on its instrumentation, which should cover the whole runtime of the application and provides a specific tuned-parameters configuration for each of the instrumented regions. The instrumentation can be inserted manually into the source code or automatically into the binary. Since we usually instrument any region for which we are able to get at least 100 power samples to have reliable energy consumption measurement (e.g., for RAPL,  it means minimum region size of 100\,ms), neither manual nor automatic instrumentation probably is providing so fine-grained annotation as the communication phases of the application are.

Both tools have active stand-alone development. The challenge is to implement a communication layer that should allow synergy of these tools, which guarantees that both tools keep the system's knowledge of which tool is in charge and what the current and future hardware settings are, without creating a conflict. This is a work-in-progress idea.

\section{Research Opportunity and Challenges}

In this section, we identify the important unsolved challenges in collective \at of two or more management layers (or domains) in the \ps. 
Identifying such challenges will provide a platform for collaborative research on \ct of different layers in the \ps. 
For \ct resource manager and runtime system, the objective is to define interactions (static and dynamic) based on the type of job (moldable/malleable vs. non-moldable/malleable) and job-agnostic interactions that are common to these two types of jobs. For co-tuning the resource manager and application, the objective is to target how the application can fully utilize the allocated resources for efficient execution while complying with the job-level power budget. For \ct runtime system and application, the objective is to fine-tune system parameters, software, and application parameters at runtime under the job-level power budget. For \ct resource manager, runtime system, and application, the objective is to explore ways in which the application can fully utilize the dynamically allocated resources for efficient execution while also maximizing the job throughput under the system power limit. When we consider \ct all four layers, we have to carefully analyze the tunable parameters from each layer to identify how they impact each other across layers, in order to find the best combination for the optimal performance under a power budget. 

We identify four areas where further research effort is required to extend the end-to-end \at for the \ps: 

\subsection{Power-Aware, Adaptive Resource Allocation}
Collective tuning of \ps layers has remained  a challenging task because of the lack of interfaces for (1) translation of high-level goals into subsequent lower-level goals, (2) translation of monitored metrics at lower layers to derived metrics at higher layers, and (3) enabling of a custom configuration (e.g., resource reallocation and remapping) at the application launch and during  runtime. Developing such interfaces will enable the platform to be used to explore several research directions for the reallocation of resources to job and sub-job components. 
In this research direction,  the following research questions should be explored. 

\begin{itemize}
\item	What target metrics are important to the individual layers of the \ps? How can the target metrics be translated into metrics understood by the lower layers of the stack assuming a top-down control transfer? 
\item	What are the different approaches to quantify the potential for performance improvement while tuning resource allocation and mapping across the stack? Potential approaches include exhaustive empirical exploration, model-based estimation, and emulation. 
\item	What customization is required in the software and hardware components participating in resource reallocation? For example, for the reallocation of resources to a running job, what features must be supported by the layers of the \ps to support the flexibility? 
\item	What operating challenges would be incurred when performing resource reallocation? For example, performing job reallocation while maintaining a minimum power draw across the system may be critical in order to comply with site-specific policies and will require additional control logic to be in place at several layers of the \ps. 
\item	Can we leverage technologies such as containers and virtual machines to address some of the challenges in resource reallocation? 
\item	Which layers of the \ps can leverage job moldability and malleability? At what point in the application run should this be done? What interfaces are provided to utilize moldability and/or malleability? 
\end{itemize}

\subsection{Offline/Static Co-tuning} 
The \ps typically operates with the software components and target binaries provided to it to solve a target problem. Several software components not directly included in the formal definition of the \ps play an important role in the eventual outcome of the \ps. Such components include compiler tool chains and the optimization features provided by them, variants of commonly used libraries (e.g., implementations of MPI, OpenMP, thread-management libraries), and input decks of the target application. These indirectly affect the efficiency of the individual layers of the \ps as well as the efficiency of the science being performed on the system. We identify the following open research questions in this topic. 

\begin{itemize}
\item	Can we quantify the impact of different compiler optimization flags for one or more target metrics on the layers of the \ps and  the application? 
\item	
Can we inform the compiler tool chain about the runtime (or online) situation on the system including resource constraints, choice of runtime algorithms, and hardware characteristics while applying optimization techniques? 
\item	Can we quantify the impact of using several variants of the application dependencies (outside of the \ps) on the efficiency of the \ps? 
\item	Can we identify correlations between black-box and/or white-box characteristics of these dependencies and the efficiency metrics relevant to the \ps? 
\end{itemize}

\subsection{Online-Offline Co-tuning on Heterogeneous Platforms}
Hardware overprovisioning has been suggested as a viable approach to address the challenges associated with site-wide or cluster-level power constraints \cite{PD13}. Since more compute and storage devices exist than can be powered up at any given time on each node of the cluster on overprovisioned hardware, the problem of selecting which components to power up and how to operate them becomes challenging. With heterogeneous computing platforms on the rise to meet the demands of exascale performance, this problem becomes even more challenging. Developing solutions to tackle this problem, which are often site-specific, will require exploring the following research questions. 

\begin{itemize}
\item	How can one quantify the trade-off between the number of compute devices on the system vs. system-level efficiency? 
\item	How can one quantify the trade-off between designing software approaches that efficiently use fewer or more compute resources vs. system efficiency? 
\item	How can one correlate software-level features into efficient usage of compute resources on the system under a power constraint? 
\item	What are the existing discrepancies between the required interfaces to drive runtime management of online or offline compute devices and existing interfaces to manage those devices? 
\end{itemize}

\subsection{Cross-Stack Parameter Tuning, Including Application-Level Controls}
Several research challenges exist in fine-tuning control parameters exposed by different software components in the \ps. The key motivation behind such a tuning is that individual layers in the stack are tuned based on a local view of those layers without taking into account the impact of the tuning decisions outside of those layers. 
We identify the following research questions in this space.  

\begin{itemize}
\item	How do the existing control parameters exposed by the different layers of the \ps and the application interact with each other on the scale of power efficiency? What are the approaches to quantify that interaction and describe the potential benefits of such exploration? 
\item	How can one find the correlation between the characteristics of the allocation and management algorithms used by the layers of the \ps and how those algorithms interact with the application? Can we augment this analysis with application-level algorithms and sub-algorithmic controls exposed by the developer? 
\item	Can we extend the algorithms at different layers of the \ps to incorporate semantic information in the application (e.g., state of the molecular dynamics simulation at each time step)? 
\end{itemize}
\section{Conclusions}

The \ps community effort has demonstrated the need for a holistic software stack for power and energy management and standardized interfaces between the layers of the \ps \cite{BB20}.
In this paper, we presented a survey of existing efforts and important open challenges to enable end-to-end \at in the \ps. First, we surveyed the layer-specific tuning efforts describing the high-level objectives, the target metrics, layer-specific control parameters, and methods, and we listed the existing software components. Then, we proposed the \ps end-to-end \at framework, identified the opportunities in \ct different layers in \ps, and presented seven  use cases. We also presented our vision of the research opportunities and challenges for collective \at of two or more management layers (or domains) in the \ps. This paper takes the first step in identifying and aggregating all the R\&D challenges related to interoperation among multiple layers of the \ps. As part of future work, we invite participation in the collaborative effort to develop the end-to-end \at framework and integrate the individual research activities, in order to realize streamlined \at of all layers of the \ps.


\section*{Acknowledgments}
This work was supported in part by LDRD funding from Argonne National Laboratory, provided by the Director, Office of Science, of the U.S. Department of Energy under contract DE-AC02-06CH11357, and in part by NSF grants CCF-1801856.

\begin{thebibliography}{}
%
%

\bibitem{1} A Strawman for an HPC PowerStack, OSTI Technical Report, August 2018 

\bibitem{BB20} A. Bartolini, S. Brink, D. Cesarini, D. Ellsworth, R. Grant, S. Jana, M. Kondo, E. K. Lee, M. Maiterth, A. Marathe, T. Patki, S. Perarnau, V. Reis, M. Schulz, O. Vysocky, T. Wilde, and X. Wu, White Paper on PowerStack, June 12, 2020. https://hpcpowerstack.github.io/raitenhaslach20.html. 

\bibitem{bhalachandra2015using}  S. Bhalachandra, A. Porterfield, and J. F. Prins (2015, May). Using dynamic duty cycle modulation to improve energy efficiency in high performance computing. In 2015 IEEE International Parallel and Distributed Processing Symposium Workshop (pp. 911--918). IEEE.

 \bibitem{bhalachandra2017adaptive} S. Bhalachandra, A. Porterfield, S. L. Olivier, and J. F. Prins (2017, May). An adaptive core-specific runtime for energy efficiency. In 2017 IEEE International Parallel and Distributed Processing Symposium (IPDPS) (pp. 947--956). IEEE.

\bibitem{CNTD} D. Cesarini, A. Bartolini, P. Bonfa, C. Cavazzoni, and L. Benini. COUNTDOWN: a run-time library for performance-neutral Energy Saving in MPI Applications, in IEEE Transactions on Computers. doi: 10.1109/TC.2020.2995269.

\bibitem{CNTDSLACK} D. Cesarini, A. Bartolini, A. Borghesi, C. Cavazzoni, M. Luisier, and L. Benini. Countdown slack: a run-time library to reduce energy footprint in large-scale MPI aplications, in IEEE Transactions on Parallel and Distributed Systems 31(11), 2696--2709, 1 Nov. 2020, doi: 10.1109/TPDS.2020.3000418.

\bibitem{tum1} I.A. Comprs Urea and M. Gerndt. (2019) Towards elastic resource management, in C. Niethammer, M. Resch, W. Nagel, H. Brunst, and H. Mix (eds.), Tools for High Performance Computing 2017.  Springer, Cham, 2019.

\bibitem{redfish} Distributed Management Task Force, Inc. (DMTF), Redfish White Paper, September 4, 2018. https://www.dmtf.org/sites/ default/files/standards/documents/DSP2044\_1.0.4.pdf. 

\bibitem{5} Energy Efficient HPC Working Group, https://eehpcwg.llnl.gov/ 

\bibitem{ES17} J. Eastep, S. Sylvester, C. Cantalupo, B. Geltz, F. Ardanaz, A. Al-Rawi, K. Livingston, F. Keceli, M. Maiterth, and S. Jana, Global extensible open power manager: a vehicle for HPC community collaboration on co-designed energy management solutions, in International Supercomputing Conference: High Performance Computing (ISC 2017), LNCS, vol. 10266, 2017.

\bibitem{HYPRE02} R. D. Falgout, J. E. Jones, and U. M. Yang, hypre: {A} Library of High Performance Preconditioners, in International Conference on Computational Science, pp. 632--641, April 2002.

\bibitem{GEOPM} GEOPM: Global Extensible Open Power Manager, https://geopm.github.io

\bibitem{PTF} M. Gerndt, E. Cesar, and S. Benkner, Automatic Tuning of HPC Applications -- The Periscope Tuning Framework (PTF). Shaker Verlag ISBN 978-3-8440-3517-9 (2015)

\bibitem{GL16} R. E. Grant, M. Levenhagen, S. L. Olivier, D. DeBonis, K. T. Pedretti, and J. H. Laros III, Standardizing power monitoring and control at exascale, IEEE Computer 49(10), 38--46, Oct. 2016. doi: 10.1109/MC.2016.308 BibTeX. 

\bibitem{GR19} R. E. Grant, B. Rountree, J. Hansen, et al., High performance computing power application programming interface specification community vVersion 1.0, Technical report, November 2019, Available at https://github.com/pwrapi/powerapi\_spec/releases/ 

\bibitem{4} HPC PowerStack, https://hpcpowerstack.github.io, https://powerstack.caps.in.tum.de.

\bibitem{3} Intelligent Platform Management Interface (IPMI), IPMI Technical Resources, 
https://www.intel.com/content/www/us/en/ servers/ipmi/ipmi-technical-resources.html 

\bibitem{tum2} Invasive Computing and HPC. http://invasic.informatik.uni-erlangen.de/en/tp{\_}d3{\_}PhIII.php

\bibitem{tum3} J. John, S. Narvaez, and M. Gerndt, Invasive computing for power corridor management, in Proceedings of the ParCo 2019: International Conference on Parallel Computing, 2019. doi: 10.3233/APC200063

\bibitem{READEXbook} P. G. Kjeldsberg, R. Schöne, M. Gerndt, L. Riha, V. Kannan, K. Diethelm, M-C. Sawley, J. Zapletal, O. Vysocky, M. Kumaraswamy, and W. E. Nagel, Runtime exploitation of application dynamism for energy-efficient exascale computing, in System Scenario-based Design Principles and Applications. Springer, Cham, ISBN 978-3-030-20342-9, doi:10.1007/978-3-030-20343-6\_6

\bibitem{PACOreadex} M. Kumaraswamy, A. Chowdhury, M. Gerndt, Z. Bendifallah, O. Bouizi, L. Riha, O. Vysocky, M. Beseda, and J. Zapletal, Domain Knowledge Specification for Energy Tuning, Concurrency Computat: Pract Exper, 2017;00:1-–6, 2017. https://doi.org/10.1002/cpe.4650

\bibitem{MA18} M. Maiterth et al., Energy and power aware job scheduling and resource management: global survey — initial analysis, in 2018 IEEE International Parallel and Distributed Processing Symposium Workshops (IPDPSW), Vancouver, BC, 2018, pp. 685--693, doi: 10.1109/IPDPSW.2018.00111. 

\bibitem{AR17} A. Marathe, R. Anirudh, N. Jain, A. Bhatele, J. Thiagarajan, B. Kailkhura, J.-S. Yeom, B. Rountree, and T. Gamblin, Performance modeling under resource constraints using deep transfer learning, in Proceedings of the ACM/IEEE International Conference for High Performance Computing, Networking, Storage and Analysis, Nov. 2017

\bibitem{AP15} A. Marathe, P. Bailey, D. K. Lowenthal, B. Rountree, M. Schulz, and B. Supinski, A run-time system for power-constrained {HPC} applications, in International Supercomputing Conference, June 2015.

\bibitem{PD13} T. Patki, D. K. Lowenthal, B. Rountree, M. Schulz, and B. Supinski, Exploring hardware overprovisioning in power-constrained, high performance computing, in Proceedings of the 27th international ACM Conference on Supercomputing, pp. 173--182, June 2013.

 \bibitem{porterfield2015application} A. Porterfield, R. Fowler, S. Bhalachandra, B. Rountree, D. Deb, and R. Lewis, Application runtime variability and power optimization for exascale computers, in Proceedings of the 5th International Workshop on Runtime and Operating Systems for Supercomputers, June 2015.
 
\bibitem{IJHPCAespreso} L. Riha, M. Merta, R. Vavrik, T. Brzobohaty, A. Markopoulos, O. Meca, O. Vysocky, T. Kozubek, and V. Vondrak,  A massively parallel and memory-efficient FEM toolbox with a hybrid total FETI solver with accelerator support. The International Journal of High Performance Computing Applications, 33(4), 660–-677, 2019. https://doi.org/10.1177/1094342018798452

\bibitem{READEX} J. Schuchart, M. Gerndt, P. G. Kjeldsberg,  et al., The READEX formalism for automatic tuning for energy efficiency. Computing 99, 727-–745, 2017. https://doi.org/10.1007/s00607-016-0532-7

\bibitem{MERIC} O. Vysocky, M. Beseda, L. Riha, J. Zapletal, V. Nikl, M. Lysaght, and V. Kannan, Evaluation of the HPC applications dynamic behavior in terms of energy consumption, in P. Ivanyi, B. H. V. Topping, and G. Varady (eds.), Proceedings of the Fifth International Conference on Parallel, Distributed, Grid and Cloud Computing for Engineering, Civil-Comp Press, Stirlingshire, UK, Paper 3, 2017. doi:10.4203/ccp.111.3

\bibitem{WU16} X. Wu, V. Taylor, J. Cook, and P. Mucci, Using performance-power modeling to improve energy efficiency of HPC applications, IEEE Computer  49(10), 20--29, Oct. 2016.


\bibitem{WK20} X. Wu, M. Kruse, P. Balaprakash, B. Videau, H. Finkel, and P. Hovland, Autotuning polybench benchmarks with Clang loop optimization pragmas, Technical Report, April 16, 2020. https://github.com/ytopt-team/autotune/blob/master/Benchmarks/polybench-autotune.pdf.

\end{thebibliography}
\end{document}